\documentclass[amsmath,amssymb,floatfix,lengthcheck,preprintnumbers,prd,showpacs,superscriptaddress,twocolumn,aps,nofootinbib]{revtex4-1}
\usepackage{graphicx}
\usepackage{slashed}

\usepackage[pdftex]{color}
\usepackage[dvipsnames]{xcolor}
\usepackage{hyperref}% add hypertext capabilities
\hypersetup{
    colorlinks=true,     % false: boxed links; true: colored links
    linkcolor=blue,      % color of internal links
    citecolor=blue,      % color of links to bibliography
    filecolor=blue,      % color of file links
    urlcolor=blue        % color of external links
}

\usepackage{cleveref}
\usepackage{mathtools}
\usepackage{xspace} % So math macros can be used in text as well -- http://tex.stackexchange.com/questions/31091/space-after-latex-commands
\usepackage{bbm}
\newcommand{\beq}{\begin{equation}}
\newcommand{\eeq}{\end{equation}}

\long\def\/*#1*/{}
\definecolor{red}{rgb}{1.0, 0, 0}

\usepackage{soul} % for strikethrough \st

\newcommand{\tr}[0]{\ensuremath {\textrm {Tr}}}
\newcommand{\vev}[1]{\ensuremath {\langle #1 \rangle}}
\newcommand{\chiPT}[0]{\ensuremath {\chi {\rm PT}}\xspace}

\newcommand{\SUNf}[0]{\ensuremath{\mathrm{SU}(N_f)}\xspace}
\newcommand{\LSB}{\ensuremath{\mathcal{L}_{\text{SB}}}\xspace}
\newcommand{\Lint}{\ensuremath{\mathcal{L}_{\text{int}}}\xspace}

\begin{document}

\preprint{MIT-CTP/5311}

\title{Soft pion scattering in infrared-conformal gauge-fermion theories}

\author{Daniel C.~Hackett}
\affiliation{Center for Theoretical Physics, Massachusetts Institute of Technology, Cambridge, MA 02139, USA}
\email{dhackett@mit.edu}

\author{Ethan T. Neil}
\affiliation{Department of Physics,
        University of Colorado, Boulder, CO 80309, USA}
\email{ethan.neil@colorado.edu}

%%%%%%%%%%%%%%%%%%%%%%%%%%%%%%%%%%%%%%%%%%%%%%%%%%%%%%%%%%%%%%%%%%%%%%%%%%%%%%
\begin{abstract}

We consider the problem of soft scattering for the analogue of pion states in gauge-fermion theories which approach a conformal fixed point in the infrared limit.  Introducing a fermion mass into such a theory will explicitly break both scale invariance and chiral symmetry, leading to confinement and a spectrum of bound states.  We argue that in such a theory, the pion scattering length diverges in the limit of zero fermion mass, in sharp contrast to QCD-like theories where the chiral Lagrangian predicts a vanishing scattering length.  We demonstrate this effect both with a simple dimensional argument, and in a generalized linear sigma model which we argue can be used to describe the interactions of light scalar and pseudoscalar bound states in the soft limit of a mass-deformed infrared-conformal theory.  As a result, lattice calculations of pion scattering lengths could be a sensitive probe for infrared scale invariance in gauge-fermion theories.

\end{abstract}


\maketitle

%%%%%%%%%%%%%%%%%%%%%%%%%%%%%%%%%%%%%%%%%%%%%%%%%%%%%%%%%%%%%%%%%%%%%%%%%%%%%%

\section{{Introduction}}

With few light fermion degrees of freedom, strongly coupled gauge-fermion theories such as QCD exhibit a rich set of physical phenomena including confinement and spontaneous chiral symmetry breaking.  As additional massless fermions are added to such a theory, eventually a transition will occur to the ``conformal window'', in which the theory has an infrared fixed point (IRFP) and confinement and chiral symmetry breaking are lost.  The interacting conformal field theories which exist in the conformal window are interesting in their own right; they also appear as important ingredients in proposals for new physics beyond the Standard Model, such as composite Higgs models \cite{Contino:2010rs,Ferretti:2013kya,Panico:2015jxa}.

Studying the transition to the conformal window is an active research topic in lattice field theory \cite{DeGrand:2015zxa,Pica:2017gcb,Svetitsky:2017xqk,Witzel:2019jbe,Drach:2020qpj}, but even for SU$(3)$ gauge theory with $N_f$ Dirac fermions in the fundamental representation, the extent of the conformal window is not well established.  Diagnosing whether a given theory has an infrared-conformal phase is a difficult task, often leading to ambiguous results.  

For example, one common method is to use lattice simulations to perform scaling tests \cite{DeGrand:2009hu,Patella:2011kp,DeGrand:2011cu,Appelquist:2011dp,Aoki:2012eq,Fodor:2012et,Fodor:2012uu,Cheng:2013xha,Bergner:2016hip,Bergner:2017gzw}: varying a perturbation such as a finite box size or an explicit fermion mass and looking for the expected power-law dependence due to the influence of the IRFP.  This is straightforward in principle, but in practice distinguishing power-law from linear behavior with a limited range of numerical data is often challenging.

One aspect where we might expect to see sharp qualitative differences is in the behavior of the pion-like states $\pi$, i.e.\ the lightest pseudoscalar mesonic bound states.  In a QCD-like theory, the pions are pseudo-Nambu-Goldstone bosons associated with spontaneous chiral symmetry breaking.  In the massless limit, their Goldstone nature manifests as a shift symmetry $\pi \rightarrow \pi + c$ which ensures that only derivative interactions are allowed: schematically,
\beq
\mathcal{L}_{\text{QCD-like}} \sim \frac{1}{2} (\partial_\mu \pi)^2  + \frac{C}{\Lambda^4} (\partial_\mu \pi)^4 + \ldots
\eeq
This means that in the \emph{soft scattering} limit of a massless QCD-like theory, the interactions between pions will vanish and the theory will become effectively free. 

On the other hand, in a mass-deformed theory in the conformal window, breaking of chiral symmetry is purely explicit and spontaneous scale generation does not occur.  In this scenario, we will argue that instead of vanishing, the scattering length $\mathbbm{a}_{\pi \pi}$ and thus the soft scattering cross section $\sigma = 4\pi \mathbbm{a}_{\pi \pi}^2$ diverges in the massless limit.

\section{Pion scattering in the soft limit and IR-conformal theories}
\label{sec:dims}

We begin 
with an introduction to the basic physical scenario of introducing mass to an infrared-conformal theory.  We then compare the expected behavior of pion scattering properties in this scenario to their behavior in chiral perturbation theory, finding a qualitative difference in their dependence on the fermion mass $m$.

\subsection{Mass-deformed infrared conformal theories}

We are interested in studying the infrared physics of a mass-deformed infrared-conformal gauge-fermion theory (``mass-deformed conformal theory'' or ``MDCT'' for short.)  The basic physical picture for such a theory is as follows \cite{Miransky:1998dh,DelDebbio:2010ze,DelDebbio:2010jy,Appelquist:2011dp}: if the fermions are exactly massless, then the theory runs to an infrared fixed point with gauge coupling $g = g^\star$, with associated mass anomalous dimension $\gamma^\star = \gamma(g^\star)$.  We may introduce an ultraviolet cutoff $\Lambda_{UV}$ such that $g(\Lambda_{UV}) \approx g^\star$, i.e.\ the theory below $\Lambda_{UV}$ is approximately scale invariant.

If we now introduce an explicit fermion mass at the UV cutoff $m \equiv m(\Lambda_{UV})$, then the mass will run to larger values in the infrared, eventually reaching the scale ${\Lambda_C \approx m(\Lambda_C)}$.  The fermions decouple completely from the theory below $\Lambda_C$, leaving a pure-gauge theory which rapidly confines, so that $\Lambda_C$ may be approximately identified as an induced confinement scale.  The relationship between $m$ and $\Lambda_C$ is determined by the mass anomalous dimension:
\beq
\Lambda_C \propto m^{1/(1+\gamma^\star)}.
\eeq
Physical quantities associated with confining physics, e.g.\ hadron masses, are then expected to be of order $\Lambda_C$.  Because $\Lambda_C$ is the only dimensionful infrared scale, all dimensionful infrared quantities are necessarily proportional to $\Lambda_C$ raised to the appropriate power (up to corrections proportional to $m \ll \Lambda_C$, as discussed in \cite{Appelquist:2011dp}).
More complicated dynamics arises if only some of the fermions are given a heavy mass, giving an infrared theory with both an induced confinement scale and dynamical fermions \cite{Brower:2014dfa,Brower:2015owo,Witzel:2018gxm,Appelquist:2020xua}; we will not explore this possibility here.

Although this basic framework is predictive and useful, going beyond simple dimensional arguments requires the construction of an effective description of the low-energy degrees of freedom.
In the case of QCD, the appropriate particle degrees of freedom are readily identified as the pions, which are much lighter than all other hadrons due to their Nambu-Goldstone nature.  However, in an MDCT there is no obvious hierarchy to the spectrum since there are no pseudo-Nambu-Goldstone bosons.  Worse yet, the masses of all hadronic states in the spectrum collapse together in the chiral limit $m \rightarrow 0$.

However, as we argue in \cref{sec:model} below, we can construct a tree-level semiclassical model for the interactions of pion-like states in an MDCT, governed by the same symmetries as the MDCT itself. We find that the requirement of recovering conformal symmetry in the $m \rightarrow 0$ limit gives powerful constraints on the allowed interactions, leading to a predictive model under certain assumptions. 

\subsection{Pion scattering}

We now turn to the main problem of interest, soft elastic scattering of the $\pi$ fields.  We focus only on S-wave scattering, since scattering into partial waves with angular momentum will be suppressed at low momentum.
We further specialize to the case of ``maximal isospin'' scattering, $\pi^+ \pi^+ \rightarrow \pi^+ \pi^+$, where $\pi^+$ is an analogue of the QCD charged pion within an SU$(2)$ subgroup of the full SU$(N_f)$ vector symmetry.
(In QCD, this is the ``$I=2$'' scattering channel.)  This is the simplest $\pi$-$\pi$ scattering process to study with lattice methods due to the lack of fermion-line disconnected diagrams.  
The combination of maximal-isospin scattering and the soft limit also yields the most robust results for the model presented in \cref{sec:scatt} below.

Two-to-two particle scattering at low momenta is often described in terms of the scattering phase shifts $\delta_\ell$ and the initial three-momentum in the center of mass frame $k$.  Specifically, at small $k$ the S-wave ($\ell = 0$) $\pi$-$\pi$ scattering process can be captured by the standard effective range expansion,
\beq
k \cot \delta_0 = -\frac{1}{\mathbbm{a}_{\pi \pi}} + \frac{1}{2} \mathbbm{r}_{\pi \pi} k^2 + \mathcal{O}(k^4),
\eeq
where $\mathbbm{a}_{\pi \pi}$ and $\mathbbm{r}_{\pi \pi}$  are known as the scattering length and effective range, respectively.  This expansion is valid so long as the scattering amplitude is a smooth function at small $k$.  

In chiral perturbation theory, the formulas for these quantities at leading order in the chiral expansion are~\cite{Gasser:1983yg}
\begin{align}
\mathbbm{a}_{\pi \pi, \chiPT} &= \frac{M_\pi}{16\pi F^2} \label{eq:chiPT_a}, \\
\mathbbm{r}_{\pi \pi, \chiPT} &= \frac{48\pi F^2}{M_\pi^3} \label{eq:chiPT_r}.
\end{align}
In the soft-scattering limit $k^2 \rightarrow 0$, the scattering cross section is $\sigma = 4\pi \mathbbm{a}_{\pi \pi}^2$.  Taking the chiral limit $m \rightarrow 0$, in which $M_\pi \rightarrow 0$ but $F$ remains finite, we see that $\mathbbm{a}_{\pi \pi} \rightarrow 0$ and the scattering cross section vanishes as the theory becomes non-interacting.  As described in the introduction, this reflects the fact that as $m \rightarrow 0$, only interactions involving the derivative $\partial_\mu \pi$ are allowed, which will vanish at zero momentum. 

The physical picture is very different in the mass-deformed conformal theory.  Up to multiplicative constants, we can simply use dimensional analysis: both of the distance scales associated with $\pi$-$\pi$ scattering must be determined by the hadronic scale $\Lambda_C$.  In other words, we must have
\beq \label{eq:MDCT_a_r_dim}
\mathbbm{a}_{\pi \pi, {\rm MDCT}} \sim \mathbbm{r}_{\pi \pi, {\rm MDCT}} \sim \Lambda_C^{-1} \sim m^{-1/(1+\gamma^\star)}.
\eeq
This is sharply different from the chiral perturbation theory behavior.  In particular, we see that in the $m \rightarrow 0$ limit, the scattering length $\mathbbm{a}_{\pi \pi} \rightarrow \infty$, \emph{diverging} rather than going to zero.

This divergence occurs because there is only one scale in the theory, the fermion mass $m$, which breaks both chiral and scale symmetry. Shift symmetry cannot arise when chiral symmetry is explicitly broken so, when ${m > 0}$, non-derivative interactions lead to a nonzero soft-scattering cross section $\sigma \sim Ck^0 + \mathcal{O}(k^2)$ with $C \neq 0$. The coefficient $C$ is dimensionful, and in an MDCT its dimensions must be given in terms of the only existing infrared scale, $\Lambda_C$.  We must therefore have $C \propto \Lambda_C^{-2} \sim m^{-2/(1+\gamma^\star)}$, which diverges as $m \rightarrow 0$.

To be more concrete, we can explicitly compare the scaling with the fermion mass $m$ in the $\chiPT$ and MDCT scenarios.  In chiral perturbation theory we have ${M_\pi \sim m^{1/2}}$ and ${F \sim m^0}$, yielding
\beq
\mathbbm{a}_{\pi \pi, \chiPT} \sim m^{1/2}, \quad \mathbbm{r}_{\pi \pi, \chiPT} \sim m^{-3/2}
\eeq
showing explicitly the form under which $\mathbbm{a}_{\pi \pi} \rightarrow 0$ as $m \rightarrow 0$.  One may also compare the ratio $\mathbbm{r}_{\pi \pi} / \mathbbm{a}_{\pi \pi}$, which diverges as $m^{-2}$ in the $\chiPT$ case but is constant at leading order in the MDCT case.

In \chiPT at somewhat heavier values of the fermion mass, particularly if $N_f$ is large \cite{Gasser:1986vb,Bijnens:2009qm}, $F$ can become dominated by its next-to-leading order correction linear in $m$ while $M_\pi$ continues to follow its leading-order scaling to good approximation.  In such a mass regime, the apparent mass dependence will be $M_\pi \sim m^{1/2}$, $F \sim m^1$, yielding the scaling $\mathbbm{a}_{\pi \pi} \sim m^{-3/2}$, $\mathbbm{r}_{\pi \pi} \sim m^{1/2}$.  Thus at heavy mass, the scattering length can also appear to diverge as $m \rightarrow 0$ in the \chiPT scenario, although the specific mass dependence is different from both the light-mass \chiPT and MDCT scaling behaviors.\footnote{Of course, if the mass $m$ is scaled properly towards the limit $m \rightarrow 0$, this scaling must give way to the expected light-mass scaling predicted by $\chiPT$.  The meaning of ``taking $m \rightarrow 0$'' here is extrapolation from data exclusively obtained in the heavy-mass regime.} In particular, we can take the dimensionless combination 
\beq
M_\pi \mathbbm{a}_{\pi \pi} \sim \begin{cases}
m^1& ({\rm light\ \chiPT}), \\
m^{-1}& ({\rm heavy\ \chiPT}), \\
m^0& ({\rm MDCT}).
\end{cases}
\eeq
We emphasize that the scaling relations above assume that corrections due to finite lattice spacing and finite volume are negligible, or at least small enough not to significantly distort the pion mass or scattering properties.

Quantities like $M_\pi \mathbbm{a}_{\pi \pi}$ can show qualitative differences in behavior as a function of $m$ in different physical scenarios, providing a powerful discriminator.
However, we caution that misleading behavior can arise in lattice studies of any quantity in an MDCT directly as a function of a dimensionless ratio like $M_\pi / F$, instead of as a function of the fermion mass $m$.
In an MDCT, the same dimensional reasoning indicates that $M_\pi / F$ is constant in $m$ at leading order; any variation seen is due to subleading scaling, finite volume effects, and lattice artifacts.
Generally, any functional relations between infrared quantities may be difficult to interpret when leading scaling dominates and all massive quantities vary with $m$ together in fixed ratios. 

\begin{figure}[h]
\includegraphics[width=0.45\textwidth]{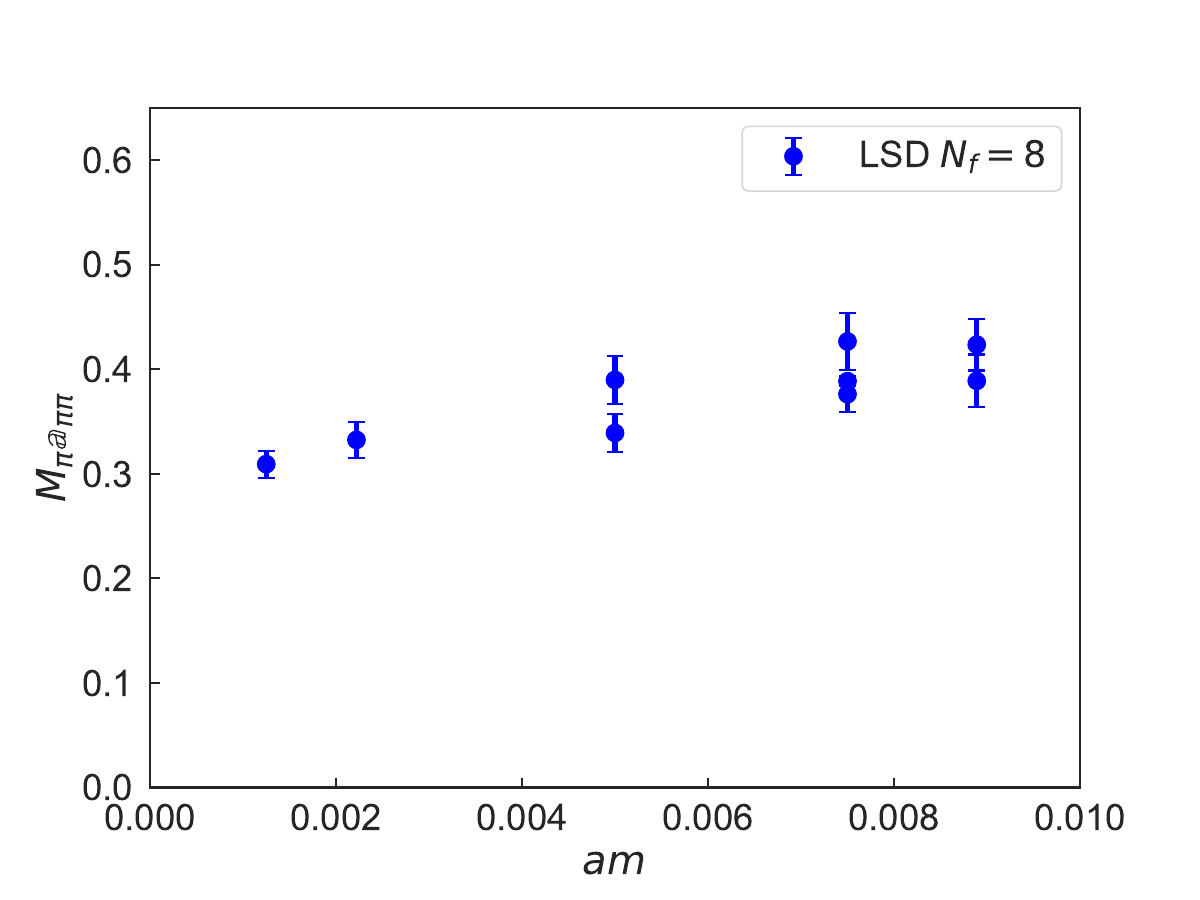}
\caption{Lattice data for the $I=2$ pion scattering length in SU$(3)$ gauge theory with $N_f = 8$ fermions in the fundamental representation, from \cite{LSD:2021xlp}.  We have re-plotted results for the combination $M_\pi \mathbbm{a}_{\pi \pi}$ against the fermion mass in lattice units $am$, where $a$ is the lattice spacing.  The relatively mild dependence on $am$ is most consistent with the low-energy description being a mass-deformed conformal or near-conformal system, as discussed in the text. \label{fig:lsd-replot}}
\end{figure}

As an example, we consider the recent lattice results from \cite{LSD:2021xlp} for $I=2$ $\pi$-$\pi$ scattering in a near-conformal theory, SU$(3)$ with $N_f = 8$ fermions in the fundamental representation.  In Fig.~\ref{fig:lsd-replot}, we re-plot their tabulated results against the fermion mass in lattice units $am$.  The leading dependence on the mass is clearly most consistent with the $m^0$ scaling of the MDCT case, and does not obviously show the scaling expected in $\chi$PT in either the light-mass or heavy-mass regimes.%
\footnote{We are assuming weak dependence of the lattice spacing on the mass in this discussion.  The lattice data shown are from staggered fermion simulations at a single bare coupling, so there is no hidden additional dependence of $a$ on other parameters.}  However, this does not conflict with the hypothesis of a dilaton effective theory considered in the reference.  Using their formula Eq.~(31) for the scattering length in the dilaton EFT, together with the mass dependences $M_\pi^2 \sim m$, $F_\pi^2 \sim m$ observed in \cite{Appelquist:2017vyy}, we find that also in the dilaton EFT case the leading behavior is $M_\pi \mathbbm{a}_{\pi \pi} \sim m^0$. We emphasize that this is an empirical argument valid only for this specific theory; in general, the leading mass scaling in the dilaton EFT scenario likely depends on the value of the mass anomalous dimension $\gamma^\star$ and on the form of the dilaton potential.

\section{Scalar infrared model for mass-deformed conformal theories \label{sec:model}}

The dimensional argument of the previous section is robust, but does not provide predictions beyond scaling relations. 
As discussed above, near-conformal systems can demonstrate similar scaling to what dimensional analysis predicts for an MDCT, making it difficult to discriminate between these scenarios on scaling arguments alone.
In this section we construct a model to study the physics of soft pion interactions in a more quantitative way.

The ideal approach would be to construct an EFT based on the weakly broken chiral and scale symmetries of the theory, then compute in the framework of that effective theory.
However, when $m \rightarrow 0$ such that scale symmetry is restored, the UV theory recovers conformality and the spectrum collapses.
This presents a formal obstacle: a collapsed spectrum cannot be cut off, so the standard procedure of constructing an invariant Lagrangian then implementing explicit breaking with spurions cannot be applied.

We instead settle for building a semiclassical tree-level model for the low-lying scalars and pseudoscalars of the MDCT with the same symmetries as the UV theory.
The construction proceeds similarly to that of an EFT, but 
does not provide a controlled approximation.
We will treat weakly broken scale invariance by constructing a theory which recovers classical conformality as $m \rightarrow 0$ and by \emph{assuming} that $m$ scales with its anomalous dimension, so our model is only applicable at tree level.

We make two assumptions about the physics of the UV theory: that the pseudoscalar mesons $\pi$, as well as the flavor singlet and non-singlet scalars $\sigma$ and $a$, are still among the lightest states in the spectrum, and that there is a spectral gap above the lightest scalar and pseudoscalar mesons.  These assumptions are supported by mass inequalities which ensure the $\pi$ are the lightest flavor-nonsinglet mesons \cite{Weingarten:1983uj,Witten:1983ut,Nussinov:1983hb,Detmold:2014iha}, as well as by lattice spectroscopy for various theories which are candidates for infrared-conformal behavior \cite{Hietanen:2008mr,Bursa:2011ru,Fodor:2011tu,Aoki:2012eq,Giedt:2012rj,Cheng:2013xha,Athenodorou:2014eua,DelDebbio:2015byq,Leino:2018yfd}.  Note that the spectral gap will vanish in absolute terms as $m \rightarrow 0$, but the relative size of the gap (i.e. mass ratios between the lightest mesons that we keep and the heavier ones integrated out of the theory) will remain constant as a result of mass hyperscaling.

To simplify the modeling task, we restrict our consideration to a particular physical process, soft $\pi$-$\pi$ scattering in the ``maximal isospin'' channel, i.e.\ soft scattering of the $\pi$ mesons in the channel with highest weight under the flavor symmetry. 
For a tree-level model, these two conditions will greatly restrict any possible contribution from intermediate states of higher spin in particular, 
meaning momentum-independent predictions of our model rely only on the weaker condition of a gap in the spectrum of spin-zero states, rather than a gap in the overall spectrum.
Given a gapped scalar spectrum, working near the soft limit (i.e.~holding the scattering momentum $k$ small compared to $\Lambda_C$) will allow us to ignore corrections from heavier scalars.

\subsection{Building the model}

We describe the $2N_f^2$ lowest-lying scalar and pseudoscalar composite meson-like states\footnote{Note that this spurion construction in terms of fermion fields does not preclude the possibility of mixing with purely gluonic states with identical quantum numbers.  The $\sigma$ state in particular has been observed to have significant overlap with a 0++ glueball state in lattice studies of near-conformal theories \cite{LatKMI:2013bhp}.} using a single complex $N_f \times N_f$ matrix scalar field  $\Phi = \Phi^i_j \sim \vev{ \overline{q}^i_R q_{Lj}}$.  
The most general Lagrangian for these modes consistent with chiral and scale symmetries is a generalized linear sigma model,
\beq
\mathcal{L} =  \tr \left[ \partial_\mu \Phi^\dagger \partial^\mu \Phi \right] + \mathcal{L}_{\rm SB} + \mathcal{L}_{\rm int}
\eeq
where $\mathcal{L}_{\rm SB}$ denotes the symmetry-breaking part of the Lagrangian which vanishes in the limit $m \rightarrow 0$, and
\begin{equation}
    \mathcal{L}_{\rm int} = -\frac{1}{4} u \tr \left[ \Phi^\dagger \Phi \right]^2 - \frac{1}{4} v \tr \left[ (\Phi^\dagger \Phi)^2 \right].
\end{equation}
In the limit $m \rightarrow 0$, the Lagrangian is classically conformal because all couplings in $\mathcal{L}_{\rm int}$ are dimensionless, and invariant under the full $\SUNf_L \times \SUNf_R$ chiral symmetry, under which the scalar fields transform as
\beq
\Phi \rightarrow U_R^\dagger \Phi U_L.
\eeq

As typically done in chiral perturbation theory, to determine the form of \LSB we introduce a spurion field $\chi$ which transforms as a scalar current under the chiral symmetry: $\chi \rightarrow U_R^\dagger \chi U_L$.  Chiral symmetry dictates how $\chi$ may appear in the chirally symmetric Lagrangian.  Setting $\chi = m \mathbbm{1}$ once the Lagrangian is constructed will explicitly break the chiral symmetry to the subgroup $\SUNf_V$ and yield the dependence of the theory on the fermion mass $m$.

In this case, we impose an additional restriction on $\chi$: it should act as a spurion field for scale symmetry as well. This is motivated by the physics of the MDCT, where we know that the fermion mass $m$ is responsible for breaking both chiral and scale symmetry.  Since the mass $m$ rescales nontrivially under dilatation, this means that any terms in $\mathcal{L}_{\rm SB}$ will include a factor of the form $\tr \left[ \chi^\dagger \chi \right]^p$ with $p$ chosen to ``correct'' the scaling of the associated operator.\footnote{Contributions of the form $\chi^p$ inside of traces cannot be included because they are not invariant under chiral symmetry.} 

Absent other constraints, an infinite number of operators may be included in \LSB.
However, as stated before, we require that $\LSB \rightarrow 0$ as $m \rightarrow 0$ so that our model recovers classical scale invariance when the explicit breaking is removed.
We impose one further physically motivated constraint: 
\beq
\lim_{\chi \rightarrow 0} \frac{\partial \mathcal{L}_{\rm SB}}{\partial \chi} = \lim_{\chi \rightarrow 0} \langle \bar{\psi} \psi \rangle = 0,
\eeq
i.e. the chiral condensate (an order parameter for chiral and scale symmetry breaking) should also vanish at ${m = 0}$.

We note in passing that for any terms in $\mathcal{L}_{\rm SB}$ that depend on non-integer powers of $\tr \left[ \chi^\dagger \chi \right]$, higher derivatives of the form $\partial^n \mathcal{L}_{\rm SB}/\partial \chi^n$ will necessarily be divergent as $\chi \rightarrow 0$ for $n$ sufficiently large.  However, in the underlying MDCT such a derivative corresponds to $n$ insertions of the $(\bar{\psi} \psi)$ operator at the same space-time point.  In the $m \rightarrow 0$ limit, we expect power-law dependence on the separation between operators in the underlying conformal field theory, which diverges if evaluated at the same space-time point.

With these conditions, and using the fact that under a passive coordinate transformation $x \rightarrow e^\lambda x$ the mass transforms as ${m \rightarrow e^{(1+\gamma_\star) \lambda} m}$, we find the form (up to redundancies which do not affect our physics of interest, discussed below) for the symmetry-breaking part of the Lagrangian:
\begin{align}
\mathcal{L}_{\rm SB} &= \tilde{B_1} \tr \left[ \chi^\dagger \chi\right]^{(3-y)/{(2y)}}  \tr \left[ \chi^\dagger \Phi + \Phi^\dagger \chi \right] \nonumber \\
&- \tilde{B_2} \tr \left[ \chi^\dagger \chi \right]^{1/y} \tr[\Phi^\dagger \Phi] \nonumber \\
&- \tilde{B_3} \tr \left[ \chi^\dagger \chi \right]^{(1-y)/y} \tr[(\chi^\dagger \Phi)^2 + (\Phi^\dagger \chi)^2] \nonumber \\
&- \tilde{B_4} \tr \left[ \chi^\dagger \chi \right]^{(1-y)/y} \left|\tr[\chi^\dagger \Phi] \right|^2 \nonumber \\
&- \tilde{B_5} \tr \left[ \chi^\dagger \chi \right]^{(1-y)/y} (\tr[\chi^\dagger \Phi]^2 + \tr[\chi \Phi^\dagger]^2)
\label{eqn:Lsb-tilded}
\end{align}
where $y \equiv 1 + \gamma^\star$ is the full scaling dimension of $m$.  We observe that the trace structure of $\tilde{B_3}$ and $\tilde{B_4},\tilde{B_5}$ is similar to the next-to-leading order terms $L_8$ and $L_6,L_7$ in the usual chiral Lagrangian \cite{Gasser:1983yg}.
Because there is no limit of this theory in which $m$ (and thus $\chi$) will scale without an anomalous dimension, the couplings $\tilde{B}_{1-5}$ are dimensionless.\footnote{Renormalization of the low-energy theory itself would induce additional anomalous dimensions on these couplings, but our model is explicitly tree-level and does not account for quantum corrections.}

There are, of course, other states in the spectrum of an MDCT that interact with the low-lying scalars, and our model should include them if these interactions are important to the physics.
We can divide these states into heavier scalars and higher-spin states; we first consider the latter.
In this work we restrict our consideration to one particular physical process, soft $\pi$-$\pi$ scattering in the maximal isospin channel.
The requirement of maximal isospin prevents any intermediate state from contributing in the $s$-channel at tree level.  This amounts to a statement of flavor number conservation: none of the quarks in the initial ``$(u\bar{d})^2$'' state may annihilate, so there is always more than one composite particle in the intermediate state.  In the $t$ and $u$ channels, the coupling of any higher-spin state to a pair of scalar particles must involve at least one derivative by Lorentz symmetry; the resulting (tree-level) scattering amplitude will be proportional to $t$ or $u$ and therefore vanish in the soft scattering limit.  We conclude that the contributions of higher-spin states can be safely neglected for study of our process of interest {at leading order, $\mathcal{O}(k^0)$, although neglecting them at $\mathcal{O}(k^2)$ requires the further assumption of a spectral gap.}

On the other hand, heavier scalar modes may contribute to soft $\pi$-$\pi$ scattering without this derivative suppression.
As stated above, we assume there is a gap in the spectrum of scalar and pseudoscalar modes above the lowest-lying states.
In the soft-scattering limit where $k \ll \Lambda_C$, this assumed gap means that heavier scalar modes will contribute as a subleading effect and we may choose to simply ignore them.

To be more quantitative about the approximation of neglecting heavier scalars, we can in principle treat their effects by extending the model to include them, adding additional fields and constructing an extended Lagrangian by applying the same constraints to the corresponding new terms in \Lint and \LSB.
After setting $\chi = m \mathbbm{1}$, we tune the couplings of the resulting model Lagrangian to reproduce the observed scalar spectrum.
We may then insert a cutoff $\Lambda$ in the gap in the spectrum \emph{of the model} and integrate these additional modes out.
The leading-order corrections are absorbed into a redefinition of the couplings $u$, $v$, and $\tilde{B}_{1-5}$.
Because the cutoff $\Lambda$ breaks scale symmetry explicitly, this process also induces an infinite number of new, previously forbidden terms suppressed by powers of $1/\Lambda$.
These operators may contribute to the tree-level scattering of process of interest after re-expanding around the vacuum; choosing to neglect them amounts to working at leading order in $M_S/\Lambda$ and $k/\Lambda$, where $S$ denotes any of the scalar or pseudoscalar states in our theory. 
These corrections are small so long as we work near the soft limit and there is a sufficiently large gap in the scalar spectrum.  The gap will shrink in absolute magnitude as $m \rightarrow 0$ and all states in the MDCT spectrum collapse together; we may still integrate out the heavier scalars as long as we adjust the cutoff $\Lambda$ downwards with the mass as well, holding the ratio $M_S/ \Lambda$ fixed.  In either case, the Lagrangian we compute with is the same, up to how we interpret the couplings.

Taking advantage of $\chi \sim m \mathbbm{1}$, we may rewrite the symmetry-breaking terms in a more convenient form
\begin{equation}
\begin{split}
	\mathcal{L}_{\rm SB}
	&= B_1 \tr \left[ \Phi + \Phi^\dagger \right] - B_2 \tr \left[ \Phi^\dagger \Phi \right]  \\
	&- B_3 \tr \left[ \Phi^2 + \Phi^\dagger{}^2 \right] - B_4 \tr \left[ \Phi \right] \tr \left[ \Phi^\dagger \right]  \\
	&- B_5 (\tr \left[ \Phi \right]^2 + \tr \left[ \Phi^\dagger \right]^2)
\end{split}
\label{eqn:LSB-dimful}
\end{equation}
where the definitions of the dimensionful $B_i$ may be read off by comparison with \cref{eqn:Lsb-tilded}.  
The potential defined by these couplings and \Lint may have multiple minima;
however, this does not correspond to spontaneous symmetry breaking in the UV theory because \LSB and thus $\vev{\Phi}$ vanishes with $m$.   

Note that from above the scaling of the dimensionful $B_i$ couplings with $m$ is
\beq
B_1 \propto m^{3/y},\ \ B_{2-5} \propto m^{2/y}.
\eeq
This scaling behavior matches on to what one would infer from hyperscaling arguments using engineering dimensions, i.e. identifying $B_1 \sim \Lambda_C^3$ and $B_{2-5} \sim \Lambda_C^2$.  Thus, our spurionic construction has successfully reproduced the desired mass hyperscaling.

In Eq.~\ref{eqn:Lsb-tilded}, we have chosen to write each term with a prefactor $\sim \tilde{B} \, (\tr [\chi^\dagger \chi] )^p$.
However, in full generality, we should write each prefactor as a sum over all operators constructed from $\chi$ consistent with chiral symmetry and scale invariance.
There are an infinite family of such operators, constructed from products of $\tr [ (\chi^\dagger \chi)^n]$ and $(\det \chi^n + (\text{c.c.}))$ (with $n \in \mathbb{Z}^+$) with each factor in the operator raised to the appropriate power to make \LSB scale invariant.
Each operator in each prefactor is accompanied by an independent dimensionless coupling.
However, this ambiguity is only in the $\Phi$-independent prefactors; the $\Phi$-dependent parts of the terms in \LSB are the only permitted structures.
As a consequence, when making the replacement $\chi \rightarrow m \mathbbm{1}$, all terms in \LSB will collapse to the simple structure of Eq.~\ref{eqn:LSB-dimful}, leaving a finite number of dimensionful couplings.  Further study of physical quantities that depend on $\chi$ beyond the leading-order replacement with its VEV, e.g.\ calculation of scalar form factors, may be able to distinguish between the additional possible structures in $\chi$.  This ambiguity has no effect on pion scattering, so we do not pursue the question further here.

We stress that scale invariance is a strong constraint on $\mathcal{L}_{\rm SB}$: {after absorbing redundancies into dimensionful couplings and} along with the requirement that $\mathcal{L}_{\rm SB}$ vanishes as $m \rightarrow 0$, we find that $\mathcal{L}_{\rm SB}$ admits only these five operators when $N_f \ge 4$.  There are additional symmetry-breaking terms including determinants which may be added to $\mathcal{L}_{\rm SB}$ for smaller values of $N_f$, for example the operator
\beq
\tr \left[ \chi^\dagger \chi \right]^{(4-N_f)/2y}  \left( \det \Phi + \det \Phi^\dagger \right),
\eeq
which vanishes in the limit $m \rightarrow 0$ so long as $N_f < 4$.  Such determinant operators cannot contribute to tree-level $\pi$-$\pi$ scattering for $N_f \geq 2$, so we do not consider them further in this work.

So far, this construction has been for the continuum limit of an MDCT, but any numerical study is necessarily carried out at finite lattice spacing $a$. Introducing the new scale $a$ into the theory will allow for additional symmetry-breaking operators.  We defer any detailed study of such lattice-dependent operators to future work, and present continuum formulas that are applicable so long as lattice artifacts are small.

\subsection{Determining the VEV and particle masses}

Setting $\chi = m \mathbbm{1}$, the explicit symmetry breaking due to \LSB will induce a VEV $\vev \Phi = (F/2) \mathbbm{1}$.  This $F$ is readily identified as the pion decay constant $F_\pi$ defined in the usual way in terms of an external axial vector current.  For comparison with chiral perturbation theory, note that our $F$ corresponds to $F_\pi$ in the 93~MeV convention.  We work with individual fields embedded in $\Phi$ using the standard non-linear representation
\beq
\Phi = \Sigma \exp \left[ i \frac{\sqrt{2}}{F} \Pi \right]
\eeq
where $\Sigma$ and $\Pi$ are Hermitian scalar and pseudoscalar matrix fields, respectively.  These fields may be further split into trace and traceless modes,
\beq
\begin{split}
\Sigma &= \frac{1}{\sqrt{2}} \left( \frac{\sigma}{\sqrt{N_f}} \mathbbm{1} + a^a X^a \right) + \frac{F}{2} \mathbbm{1}, \\
\Pi &=  \frac{\eta'}{\sqrt{N_f}} \mathbbm{1} + \pi^a X^a,
\end{split}
\eeq
where $X^a$ are the generators of the $\mathfrak{su}(N_f)$ algebra, normalized such that $\tr [X^a X^b] = \delta^{ab}$.
For convenience, we define the matrix fields $a = a^a X^a$ and $\pi = \pi^a X^a$.  The various factors accompanying the fields give canonically normalized kinetic terms for the (real) $\sigma^a$, $a^a$, and $\pi^a$ fields.  We expand around the VEV by including the term $\sim F$ in $\Sigma$.  The $\eta'$ mode is generally much heavier than the other states due to the $U(1)_A$ anomaly; we integrate it out ``by hand'' by simply setting $\eta' = 0$, as in \cite{Appelquist:2018tyt}. (This amounts to replacing $\Pi \rightarrow \pi$.)

We rewrite our Lagrangian in terms of the $\sigma$, $a$, and $\pi$ fields and only retain terms relevant to tree-level $\pi$-$\pi$ scattering, obtaining our working Lagrangian ${\mathcal{L}_{W} = \mathcal{L}_1 + \mathcal{L}_2 + \mathcal{L}_3 + \mathcal{L}_4}$.
We retain interactions with derivatives arising from the $\Phi$ kinetic term, which give nonvanishing contributions to $s$ channel processes.
The resulting noninteracting part of the Lagrangian is
\begin{equation} \begin{split}
\mathcal{L}_2 &=
\frac{1}{2} (\partial_\mu \sigma)^2
+ \frac{1}{2} \tr (\partial_\mu a)^2
+ \frac{1}{2} \tr (\partial_\mu \pi)^2
\\ &
- \frac{1}{2} M_\sigma^2 \sigma^2
- \frac{1}{2} M_{a}^2 \tr \, a^2
- \frac{1}{2} M_\pi^2 \tr \, \pi^2,
\end{split} \end{equation}
with
\begin{equation} \begin{split}
M_\sigma^2 &= B_2  + 2B_3 + N_f B_{45} + \frac{3F^2}{8} (N_f u + v), \\
M_{a}^2 &= B_2 + 2B_3 + \frac{F^2}{8}  (N_f u + 3 v) \\
&= M_\sigma^2 - \frac{1}{4} N_f u F^2 - N_f B_{45}, \\
M_\pi^2 &= \frac{2B_1}{F} - 4B_3 - N_f B_{45}, 
\end{split} \end{equation}
where the two couplings $B_4$ and $B_5$ always occur in the characteristic combination 
\begin{equation}
B_{45} \equiv B_4 + 2B_5.
\end{equation}
The three-point part of the Lagrangian pertinent to tree-level $\pi$-$\pi$ scattering is
\begin{equation} \begin{split}
\mathcal{L}_3 &= g_{a\pi} \tr \left[ a \pi^2 \right] + g_{a\partial} \tr \left[ a (\partial_\mu \pi)^2 \right] 
\\ &
+ g_{\sigma \pi} \sigma \tr \left[ \pi^2 \right] 
% \\ &
+ g_{\sigma \partial} \sigma \tr \left[ (\partial_\mu \pi)^2 \right],
\end{split} \end{equation}
where the couplings are given by
\begin{equation}\begin{split}
g_{a\pi} &= -\frac{1}{\sqrt{2} F} \left[ M_\pi^2 - 4B_3 \right], \\
g_{a\partial} &= \frac{\sqrt{2}}{F}, \\
g_{\sigma \pi} &= -\frac{1}{\sqrt{2N_f} F} \left[ M_\pi^2 - 4B_3 - N_f B_{45} \right], \\
g_{\sigma \partial} &= \sqrt{\frac{2}{N_f}} \frac{1}{F}.
\end{split}\end{equation}
The pertinent four-point part of the Lagrangian is
\begin{equation} \begin{split}
\mathcal{L}_4 &= 
\frac{1}{6 F^2} \tr \left[
	(\pi \partial_\mu \pi)^2 - \pi^2 (\partial_\mu \pi)^2 \right] \\
	&+ \frac{1}{12F^2} \left[  (M_\pi^2 - 12B_3) \tr \left[ \pi^4 \right] - 3B_{45} \tr \left[ \pi^2 \right]^2 \right].
\end{split} \end{equation}
and the one-point coupling for the $\sigma$ is
\begin{equation} \begin{split}
\mathcal{L}_1 = \sqrt{\frac{N_f}{2}} & \left[ 
\vphantom{\frac{1}{8}}
2 B_1 - F B_2 - 4F B_3 - F N_f B_{45} \right. \\
& \left. - \frac{1}{8} F^3 (N_f u + v) \right] \sigma
.
\end{split}
\label{eqn:b-def}
\end{equation}
Solving $\vev{\Sigma} = \frac{F}{2} \mathbbm{1}$ using the one-point coupling for $\sigma$ gives a cubic equation for $F$ in terms of the couplings in the original Lagrangian (neglecting $\Lambda$-suppressed contributions from \Lint, if these are included). The full solution for $F$ is complicated and unenlightening, but it is straightforward to verify based on the mass dependence of the parameters $B_i$ that $F \propto m^{1/y}$ as expected for a hadronic scale. 
As discussed above, because $\LSB \rightarrow 0$ as $m \rightarrow 0$, we do not require the system to be in any particular vacuum or make any further restrictions on the parameters from the potential.

Note that the interactions of pions arise entirely from \LSB and the kinetic term for $\Phi$, and the couplings $u$ and $v$ from $\mathcal{L}_{\text{int}}$ appear only in the masses.
This is because the operators associated with $u$ and $v$ in \Lint do not involve derivatives and thus, due to chiral symmetry, only involve the scalar fields $\Sigma$.

% \vfill\null 
% \newpage

For the calculations to follow, we will trade the set of Lagrangian couplings $\{u, v, B_1, B_2\}$ for the more physical set $\{M_\pi, M_\sigma, M_{a}, F\}$, all of which scale as $m^{1/y}$ as expected from dimensional analysis.
The dependence on $B_3$ and $B_{45}$ unavoidably remains; removing them in favor of more physical couplings would require additional measurements to match.

\vfill\null

\newpage

\subsection{Scattering lengths of pseudoscalar mesons}
\label{sec:scatt}

Determining the scattering length and effective range in a given theory requires calculating the Feynman amplitude $\mathcal{M}$ for the process of interest, performing a partial-wave expansion of the amplitude, converting the partial-wave amplitude to a phase shift by including kinematic factors, and finally expanding at small $k$.  There are multiple conventions in the literature for carrying out this procedure in full; we attempt to gather and reconcile the conventions we are aware of in Appendix~\ref{app:finiterange}.  

\begin{widetext}
We find the tree-level maximal-isospin $\pi$-$\pi$ scattering amplitude to be
\begin{equation}
\begin{split}
\mathcal{M}_{\rm MDCT}(s,t,u) 
&= 2 \frac{M_\pi^2}{F^2} \left[1 - \frac{s}{2M_\pi^2} - \frac{4B_3 + 2B_{45}}{M_\pi^2} \right]
\\
&
- \left[ 2\left(1 - \frac{2}{N_f} \right) \frac{[M_\pi^2 + 4B_3 - t ]^2}{t-M_a^2} 
+ 4 \frac{[M_\pi^2 + 4B_3 + N_f B_{45} - t]^2}{t-M_\sigma^2} + (t \leftrightarrow u) \right].
\end{split}
\label{eq:M-mdct}
\end{equation}
In the limit where $M_a, M_\sigma \rightarrow \infty$ and  $B_3 = B_{45} = 0$, our theory becomes identical to the leading-order $\chiPT$ Lagrangian.  Taking this limit can thus provide a simple check of our results.  We find the result
\beq
\lim_{M_a, M_\sigma \rightarrow \infty} \left. \mathcal{M}_{\rm MDCT}(s,t,u) \right|_{B_3, B_{45} = 0} = \frac{M_\pi^2}{F^2} \left(2 - \frac{s}{M_\pi^2} \right) 
\eeq
which exactly matches the amplitude%
\footnote{Note that the maximal-isospin scattering amplitude in the representation theory of the reference is denoted as ``SS''.}  obtained in \cite{Bijnens:2011fm} 
at leading order for the same chiral symmetry breaking pattern, $\SUNf \times \SUNf \rightarrow \SUNf$.
If this result holds, the caveats discussed at the end of \cref{sec:dims} especially apply about studying $\mathbbm{a}_{\pi \pi}$ as a function of $M_\pi / F$ (which is identical $\chiPT$ and MDCT cases), instead of as a function of the fermion mass $m$ (which will show strongly different behavior in the two cases).
Given the expected lightness of the dilatonic $\sigma$ in an MDCT, and the absence of any reason for $B_3$ and $B_{45}$ to be small, this limit is unlikely to be a good description of the MDCT on physical grounds.

Going back to the full amplitude and taking the limit as $k^2 \rightarrow 0$, we obtain%
\footnote{Note that $\cos \theta$ dependence appears at $O(k^4)$. One may check that computing $k \cot \delta_0$ from Eq.~\ref{eq:M-mdct} and then expanding in small $k^2$ yields the same results.} 
\begin{equation}
\mathcal{M}_{\rm MDCT}(s,t,u) = -2 \frac{M_\pi^2}{F^2} \left[1 + \frac{4B_3 + 2B_{45}}{M_\pi^2} \right]  + 2 \frac{N_f - 2}{N_f} \frac{(M_\pi^2 + 4B_3)^2}{M_a^2 F^2} + \frac{4}{N_f} \frac{(M_\pi^2 + 4B_3 + N_f B_{45})^2}{M_\sigma^2 F^2} + \mathcal{O}(k^2).
\end{equation}
Expanding in $k^2$ as described in Appendix~\ref{app:finiterange}, this leads to
\beq \label{eq:MDCT_a}
\mathbbm{a}_{\pi \pi, {\rm MDCT}} = \frac{1}{16\pi} \frac{M_\pi}{F^2} \left[
1 + \frac{4B_3 + 2B_{45}}{M_\pi^2} 
- \frac{N_f - 2}{N_f M_\pi^2 M_a^2} (M_\pi^2 + 4B_3)^2 
- \frac{2}{N_f M_\pi^2 M_\sigma^2} (M_\pi^2 + 4B_3 + N_f B_{45})^2 \right]
.
\eeq
\end{widetext}
The expression for $\mathbbm{r}_{\pi \pi, {\rm MDCT}}$ is more complicated and its validity relies on the stronger assumption of a gap in the overall spectrum, so we do not provide it here.
We emphasize again that this is a tree-level result.  We see that the corrections to the scattering length due to the presence of the scalars (the last two terms) are manifestly negative, regardless of the signs of the individual couplings.

A possible intermediate scenario is that some of the scalars included in the theory are relatively heavy, and a more consistent description is given by integrating them out.  For example, lattice spectroscopy for SU$(3)$ with $N_f = 12$ fermions (summarized in \cite{Brower:2015owo}) indicates that ${M_\pi, M_\sigma < M_a \sim M_\rho}$, where $\rho$ is the lightest vector meson.  We can obtain a simplified tree-level result with the flavored $a$ scalars removed by taking $M_{a} \rightarrow \infty$ in our formulas above.  In this limit, we find
\begin{equation}\begin{split}
\lim_{M_a \rightarrow \infty} &\mathbbm{a}_{\pi \pi, {\rm MDCT}} 
= \frac{M_\pi}{16\pi F^2} \left[ 1 + \frac{4B_3 + 2B_{45}}{M_\pi^2} 
% \right. \nonumber \\ &\left. 
\right. \\ &\left.
- \frac{2}{N_f M_\pi^2 M_\sigma^2} (M_\pi^2 + 4B_3 + N_f B_{45})^2 \right]. \label{eq:MDCT_limit_a} 
\end{split}\end{equation}

It is straightforward to verify that both the simplified heavy-scalar expression \cref{eq:MDCT_limit_a} and the full tree-level result \cref{eq:MDCT_a} (as well as the corresponding expressions for $\mathbbm{r}_{\pi \pi, {\rm MDCT}}$ which are not shown explicitly) 
yield precisely the mass dependence that was argued for on dimensional grounds in \cref{eq:MDCT_a_r_dim}.  

In both the full tree-level result \cref{eq:MDCT_a} and the simplified heavy-scalar expression \cref{eq:MDCT_limit_a}, we find different leading-order dependence of $\mathbbm{a}_{\pi \pi}$ on $M_\pi$ than in chiral perturbation theory.
Additionally, unlike in \chiPT, we find leading-order dependence on $M_\sigma$ and $M_a$.
It may be possible to differentiate between chiral and conformal scenarios by examining this dependence, especially if $B_3$ and $B_{45}$ can be constrained using independent physical measurements.
However, we note that this model does not treat subleading scalings, finite volume effects, or lattice artifacts. 
We refer again to the caveats discussed at the end of \cref{sec:dims} about difficulties using discriminators based on functional relationships between infrared quantities and the related problems with studying quantities like $\mathbbm{a}_{\pi \pi}$ as a function of $M_\pi / F$ instead of $m$.

\section{Conclusion}\label{sec:conclusion}

We have studied the scattering of pseudoscalar mesons in mass deformations of gauge-fermion theories which are conformal in the infrared limit.  Focusing on the soft scattering limit, our analysis finds that the scattering length $\mathbbm{a}_{\pi \pi}$ in an MDCT diverges as $m^{-1/y}$ in the $m \rightarrow 0$ limit, qualitatively different from chiral perturbation theory where $\mathbbm{a}_{\pi \pi} \rightarrow 0$.  Thus, using lattice methods to measure the dependence of the scattering length on the quark mass should allow these two cases to be easily distinguished.  On the contrary, studies of $M_\pi \mathbbm{a}_{\pi \pi}$ vs. $M_\pi^2 / F^2$ may give misleading conclusions as the leading dependence in both the MDCT and \chiPT cases is similar.  (The primary difference is that in the MDCT case $M_\pi^2 / F^2$ will be constant, but this may be obscured by corrections due to finite lattice spacing and volume or working at relatively heavy $m$.)

An important caveat is that for theories which are confining but just below the edge of the conformal window, the low-energy regime may be described 
by an effective theory
such as a linear sigma model \cite{Meurice:2017zng,Appelquist:2018tyt,DeFloor:2018xrp} or a dilaton EFT \cite{Matsuzaki:2013eva,Golterman:2016lsd,Golterman:2016cdd,Kasai:2016ifi,Hansen:2016fri,Appelquist:2017wcg,Appelquist:2017vyy,Golterman:2018mfm}.  In the latter case especially, it has been pointed out that the spectrum in the large-mass regime can show the same power-law scaling with $m$ as expected for an MDCT \cite{Golterman:2018mfm}, and recent results \cite{LSD:2021xlp} found that $\mathbbm{a}_{\pi \pi}$ is constant at leading order in $m$ in this regime.
Comparison of more detailed predictions of dilaton EFTs versus those of MDCT-specific models like the one presented here may allow disambiguation beyond what is possible from simple scaling arguments.

Although our key results depend mainly on dimensional arguments, we have also constructed a tree-level semiclassical model in which soft $\pi$-$\pi$ scattering can be studied, based on the strong assumption that the spectrum of the MDCT has a gap above the lowest-lying scalar and pseudoscalar modes. Even if this assumption does not hold, so long as the scalar part of the spectrum is gapped, our analysis may still be applied as a model for maximal isospin $\pi$-$\pi$ scattering in the soft limit as contributions from higher-spin states can be argued to vanish on general grounds.  It may be interesting to consider the calculation of other quantities within this low-energy theory, such as scalar and vector form factors for the light scalar states for which there are some existing predictions from scaling arguments \cite{DelDebbio:2010jy,DelDebbio:2013qta}.

Extending the model to include finite lattice spacing effects could provide a more powerful discriminator.
As discussed in Sec.~\ref{sec:dims}, hyperscaling means dimensionful quantities vary in fixed ratios near the chiral limit, complicating the use of relations between infrared quantities to discriminate between conformal and other scenarios, but corrections due to lattice artifacts could disambiguate this relationship.
However, finite lattice spacing corrections add an additional scale which may significantly complicate the construction of the model.

It may be possible that our tree-level model can be extended to include quantum corrections.
The requirement used here that, in the symmetric limit $\LSB \rightarrow 0$, the model recovers classical conformality becomes a requirement that the theory sits at a fixed point when $\LSB \rightarrow 0$.
If the scalar fields themselves can acquire large anomalous dimensions, additional operators may be allowed in the symmetry-breaking Lagrangian \cref{eqn:Lsb-tilded}.
Given these complications, it may not be possible to neglect higher-spin and heavier scalar modes as in the tree-level approach.
We leave the necessary further formal work to this end for future studies.

The formulas presented here for the low-energy model assume the chiral symmetry breaking pattern $\SUNf \times \SUNf \rightarrow \SUNf$, corresponding to fermions charged in a complex representation of the underlying gauge theory.  Different symmetry breaking patterns are known to occur if the fermion representation is instead real or pseudoreal, or in the case of mixed fermion representations; we leave the study of these possibilities to future work.  Changing the chiral symmetry breaking pattern will modify the detailed formulas for $\mathbbm{a}_{\pi \pi}$ and $\mathbbm{r}_{\pi \pi}$ but not the qualitative dependence on the fermion mass.

\begin{acknowledgements}
The authors would like to thank Tom Appelquist and Mike Buchoff for very helpful discussions in the early stages of this work.  We also thank Tom DeGrand and Ben Svetitsky for enlightening conversation and Maarten Golterman, Yigal Shamir, Andy Gasbarro, and James Ingoldby for useful comments on an early draft of the manuscript.  DCH thanks Anthony Grebe, Gurtej Kanwar, Patrick Ledwidth, and Michael Wagman for useful discussions on conformality.  ETN is supported by the U.~S.~Department of Energy, Office of Science, Office of High Energy Physics, under Award Number DE-SC0010005. DCH is supported in part by the U.S. Department of Energy, Office of Science, Office of Nuclear Physics under grant Contract Numbers DE-SC0011090 and DE-SC0021006. 
\end{acknowledgements}

%%%%%%%%%%%%
\bibliography{mdCFT}

\appendix

\section{Conventions for pion scattering amplitudes and finite range expansion \label{app:finiterange}}

Determining the scattering phase shifts begins with partial-wave expansion of the Feynman amplitude $\mathcal{M}$ for the two-particle scattering process:
\beq \label{eq:partialwave}
\mathcal{M}(k,\theta) = \mathcal{N} \sum_{\ell=0}^\infty A_\ell(k) (2\ell+1) P_\ell(\cos \theta)
\eeq
where $P_j$ is the $j$-th Legendre polynomial and the amplitude has been integrated over the azimuthal angle $\phi$.  We leave the normalization $\mathcal{N}$ arbitrary for now, to elucidate how it affects the partial wave expansion; it will cancel out in the final expression for the phase shift.  Our conventions for the Feynman amplitude $\mathcal{M}$ are such that for elastic $2 \rightarrow 2$ scattering in the CM frame, the differential cross section is
\beq \label{eq:ds_QFT}
\frac{d\sigma}{d\Omega} = \frac{1}{64\pi^2 s} |\mathcal{M}(k,\theta)|^2
\eeq
where $s = 4(k^2 + M_\pi^2)$.  

To relate the partial wave amplitudes $A_\ell(k)$ to scattering phase shifts $\delta_\ell(k)$, we match on to textbook quantum mechanical scattering theory in which the latter are defined.  The phase shifts are defined from the partial wave amplitudes $f(k,\theta)$ satisfying
\beq
f(k,\theta) \equiv \frac{1}{k} \sum_{\ell=0}^\infty (2\ell+1) \frac{e^{2i\delta_\ell} - 1}{2i} P_\ell(\cos \theta),
\eeq
where the $1/k$ comes from the identification of $f(k,\theta)$ as the amplitude of an outgoing spherical wave which dies off with the distance $r$ from the origin as $1/r$.  

With identical bosons in the final state, which is always the situation if we consider elastic scattering of isospin eigenstates, the quantum mechanical differential cross section given in terms of the partial-wave amplitudes must explicitly include an exchange symmetry, which leads to the form
\beq \label{eq:ds_QM}
\frac{d\sigma}{d\Omega} = |f(k,\theta) + f(k,\pi - \theta)|^2.
\eeq
If $f$ is independent of $\theta$, as for the S-wave, then we pick up a factor of 4 in $d\sigma/d\Omega$.  

Matching together \cref{eq:ds_QM,eq:ds_QFT}%
\footnote{Note that the QFT expression requires no modification for identical particles---that is accounted for in the amplitude already.  However, for both QFT and QM there will be an extra factor of 1/2 if we go on to compute the total cross section, since $d\Omega$ should be integrated only over half of phase space with identical particles.}, we find that they are equivalent if we make the identification
\beq \label{eq:pw_match}
A_\ell(k) = \frac{32\pi}{\mathcal{N}} \frac{\sqrt{s}}{2k} \frac{e^{2i\delta_\ell(k)}-1}{2i} = \frac{32\pi}{\mathcal{N}} \frac{\sqrt{k^2+m_\pi^2}}{k \cot \delta_\ell(k) - ik}.
\eeq
This equation for $A_\ell(k)$ and the partial wave expansion \cref{eq:partialwave} match the corresponding equations in the chiral perturbation theory literature \cite{Gasser:1983yg,Bijnens:2011fm} with the choice $\mathcal{N} = 32\pi$.

Focusing on S-wave scattering, using the orthogonality relation between the Legendre polynomials
\beq
\int_{-1}^1 P_j(\cos \theta) P_k(\cos \theta) d \cos \theta = \frac{2}{2j+1} \delta_{jk}
\eeq
we find the partial wave amplitude
\beq
A_0(k) = \frac{1}{2\mathcal{N}} \int_{-1}^1 \mathcal{M}(k, \theta) d\cos \theta.
\eeq
The phase shift in the usual form is equal to
\beq
k \cot \delta_0 = ik + \frac{32\pi \sqrt{k^2+m_\pi^2}}{\mathcal{N} A_0(k)}.
\eeq
The presence of the purely imaginary term $ik$ is somewhat odd-looking, but the optical theorem gives the relationship
\beq
{\rm Im}\ A_0 = \frac{2k}{\sqrt{s}} |A_0|^2.
\eeq
In particular, this indicates that for a tree-level calculation in perturbation theory, the imaginary part begins at second-order.  Therefore at leading order, $A_0$ will be real and we should work only with the real part of the formula:
\beq
k \cot \delta_0 = \frac{64\pi \sqrt{k^2 + m_\pi^2}}{\int_{-1}^1 \mathcal{M}_{\rm tree}(k,\theta) d\cos \theta} + \textrm{(higher-order)}
\eeq

Because it can be somewhat confusing, we emphasize once more that these formulas apply for indistinguishable particles in the final state.  If the final-state particles are distinguishable, then the quantum mechanical cross section \cref{eq:ds_QM} should be replaced with simply $|f(k,\theta)|^2$, which reduces the partial wave amplitude \cref{eq:pw_match} by a factor of 2.  This factor of 2 will reappear when converting between scattering amplitudes in the isospin basis and charged pion basis \cite{Petersen:1971ai,Cabibbo:2005ez}.  For example, the S-wave amplitude $A_0^{(0+)}(k)$ for scattering $\pi^0 \pi^+ \rightarrow \pi^0 \pi^+$ is equal to half of the isospin-2 amplitude, $A_0^{(0+)} = A_0^{(I=2)}/2$.

Expanding the quantity $k \cot \delta_0$ directly at small momentum $k$ gives the standard effective range expansion for S-wave scattering,

\beq
k \cot \delta_0 = -\frac{1}{\mathbbm{a}_{\pi \pi}} + \frac{1}{2} \mathbbm{r}_{\pi \pi} k^2 + \mathcal{O}(k^4).
\eeq

A common alternative used in chiral perturbation theory \cite{Gasser:1983yg,Bijnens:2011fm} is to series expand the partial-wave amplitude $A_\ell(k)$ itself as:
\beq
{\rm Re}\ A_\ell(k) = k^{2\ell} \left[ a_\ell + k^2 b_\ell + \mathcal{O}(k^4) \right].
\eeq
Setting $\ell=0$ and expanding out \cref{eq:pw_match} gives the result
\beq
k \cot \delta_0 = \frac{M_\pi}{a_0} + \left[\frac{a_0-2b_0M_\pi^2}{2a_0^2 M_\pi} - \frac{2a_0}{M_\pi} \right] k^2 + \mathcal{O}(k^4) .
\eeq
Comparing to the standard effective-range expansion, we find the relations
\begin{align}
\mathbbm{a}_{\pi \pi} &= -\frac{a_0}{M_\pi}, \\
\mathbbm{r}_{\pi \pi} &= \frac{a_0 - 2b_0 M_\pi^2}{a_0^2 M_\pi} - \frac{2a_0}{M_\pi}.
\end{align}
which have been noted before in the $\chiPT$ literature \cite{Bijnens:1997vq}.  These expressions can be used with the results of e.g.\ \cite{Gasser:1983yg} to reproduce the formulas \cref{eq:chiPT_a,eq:chiPT_r}.

\end{document}